\documentclass[12pt,a4paper]{article}
\usepackage{latexsym}
\usepackage{amsfonts}

\newcommand{\danger}[1]{\textbf{#1}}

\input{epsf}             
\def\epsfcenter#1{{\vcenter{\hbox{\epsfbox{#1}}}}} 

\usepackage[dvips]{graphicx}

\begin{document}

\title{\danger{Spin Foam Models of 
$n$-dimensional Quantum Gravity, and
Non-Archimedan and Non-Commutative Formulations}}
\author{\danger{J.Manuel Garc\'\i a-Islas}}

\date{}

\maketitle

\hspace{3.1cm}Centro de Investigacion en Matem\'aticas

\hspace{5.1cm} A.P. 402, 36000

\hspace{4.2cm} Guanajuato, Gto, M\'exico

\hspace{4.5cm} email: islas@cimat.mx

\vspace{1.3cm}

\danger{Abstract}. 
This paper is twofold. First of all 
a complete unified picture of $n$-dimensional 
quantum gravity is proposed in the following sense:
In spin foam models of quantum gravity
the evaluation
of spin networks play a very important role. These evaluations correspond
to amplitudes which contribute in a state sum model of quantum gravity. 
In \cite{fk}, the evaluation of spin networks as integrals
over internal spaces was described. This evaluation was restricted to 
evaluations of spin networks in $n$-dimensional Euclidean quantum gravity.
Here we propose that a similar method can be considered to include 
Lorentzian quantum gravity. We therefore describe the 
the evaluation of spin networks in the Lorentzian framework of spin foam models.
We also include a limit of the Euclidean and Lorentzian spin foam models which 
we call Newtonian. This Newtonian limit was also considered in \cite{jm}. 

Secondly, we propose an alternative
formulation of spin foam models of quantum gravity with its
corresponding
evaluation of spin networks. This alternative formulation is 
a non-archimedean or $p$-adic spin foam model. The interest on this description
is that it is based on a discrete space-time, which is the expected situation we 
might have at the Planck length; this description might lead us
to an alternative regularisation of quantum gravity. Moreover a non-commutative
formulation follows from the non-archimedean one.

\newpage

\section{\bf{Introduction}}

In \cite{jm}, a definition for the evaluation of spin networks in
3-dimensional quantum gravity was considered. This was done in 
a unified spirit of three cases of quantum gravity which are
the Lorentzian, the Euclidean and the Newtonian cases. 
\footnote{This latter case was given its name first in \cite{jm}, and
should be interpreted as a mathematical limit. This is because 
physically, there may not be sense for quantizing Newtonian gravity.
This is better expained in section 4.}  
Particular attention was paid to the evaluation and asymptotics of the tetrahedron(3-simplex)
network. Then it was suggested that the same work could be generalised
by considering evaluations
of spin networks in the $n$-dimensional case.

The evaluation of spin networks in the $n$-dimensional case was considered
and restricted to the Euclidean case in \cite{fk}. This evaluation  
was described as 
Feynman graphs. These are integrals over internal spaces.
 
In this paper we generalise the work done in \cite{jm} and at the same time
we complete in part the story of \cite{fk} by considering the
evaluation of the $n$-dimensional Lorentzian and Newtonian $n$-simplex.
We restrict ourselves to the case of the principal unitary irreducible
representations of $SO(n-1,1)$ and $ISO(n-1)$.

In section 2 we summarise the results on $n$-dimensional
Euclidean quantum gravity studied in \cite{fk} and 
give the recipe of
the evaluations of spin networks as simple graphs. 
\footnote{A formulation of the existence of higher dimensional gravity theories 
appears in \cite{fkp}, therefore it is worth to study the evaluation
of spin networks in any dimension.}

In section 3 we describe the case of n-dimensional Lorentzian quantum gravity
and consider the case of spacelike spin networks, that is, graphs whose
edges are all spacelike.

In section 4 we describe the limitng case of Newtonian spin networks,
a limit of both, Euclidean and Lorentzian quantum gravity.
This case should be interpreted as a mathematical idea which physically
is given when the
speed of light is so large. However, we doubt its physical interpretation
of a Newtonian quantum gravity as this may not have sense at all as it will be 
addressed in section 4.
Anyway, it is still interesting to have a physical interpretation
of the Newtonian spin networks and to explore its impotance
for spin foam models, if any.

The second part of the paper starts in section 5 
where we propose a study of a $p$-adic and quantum deformation
formulation of quantum gravity inspired on the spin foam models.
There is a possible relation to a non-archimedean and non-commutative 
geometry respectively. 
This last section may be of great interest although  
future work is required to give it a precise formulation and to 
develop its possible importance. There is a thought that an adelic description
of quantum gravity in terms of spin foams may be needed.

\section{Euclidean spin networks}

A way to evaluate simple spin networks for the group $SO(n)$ 
was introduced in \cite{fk}. In this section we give a description
of this idea which follows similarly for the Lorentz group $SO(n-1,1)$ and for the
inhomogeneous group $ISO(n-1)$ which will be discussed in the following sections.

Simple spin networks of the group $SO(n)$ are those constructed from special 
representations of $SO(n)$. A special class of representations are the spherical
harmonics \cite{h} that appear in the decomposition in the decomposition of the
space of fuctions $L^{2}(\mathbf{S}^{n-1})$ into irreducible components.
These representations are labelled by integers $\ell$ and the spin networks,
$\Sigma$,
have edges labelled by $\rho$'s. The evaluation of these spin networks is
given by the following rules which closely resemble the rules of Feynman graphs
that apear in quantum field theory:

\bigskip

-With every edge of the graph $\Sigma$, associate a propagator $K_{\rho}^{E}(x,y)$.

\bigskip

-Take the product of all these data and integrate over one copy of the homogeneous
space $SO(n)/SO(n-1)=\mathbf{S}^{n-1}$ for each vertex.

\bigskip

Thus the evaluation formula is given by

\begin{equation}
\int_{\mathbf{S}^{n-1}} \prod_{v} dx_{v} \prod_{e} K_{\rho_{e}}^{E}(x,y)
\end{equation}
where $\rho_{e}$ denotes the representation labelling the edge $e$.

The propagators $K_{\rho}^{E}(x,y)$ are given by the Gegenbauer polynomials $C_{n}^{p}$
as described in \cite{fk}.

In \cite{jm} we discussed that the propagators in n-dimensional quantum gravity for any
case(Euclidean, Lorentzian, Newtonian) are given by zonal spherical functions of 
the respective group. In this Euclidean case these propagators are Legendre polynomials
and are directly related to the Gegenbauer polynomials \cite{vk}.

These propagators are then expressed in an integral form as

\begin{equation}
K_{\rho}^{E}(r)= \frac{\Gamma(\frac{n-1}{2})}
{\sqrt{\pi}\Gamma(\frac{n-2}{2})}
\int_{0}^{\pi}
( \cos \theta + i \cos \psi \sin \theta)^{\sigma} \sin^{n-3} \psi d\psi
\end{equation}
where $\sigma = -p + i \ell$. $p$ is related to the dimension of the 
space-time as $p=(n-2)/2$ and $\ell$ labels the irreducible representations os
$SO(n)$.

As a special case we have the n-simplex $K^{n+1}$, which is the complete graph with 
$n+1$ vertices.\footnote{A complete graph $K^{n+1}$ has $n+1$ vertices and an edge
for each pair of vertices, so that it has $n(n+1)/2$ edges.}

\[ \epsfcenter{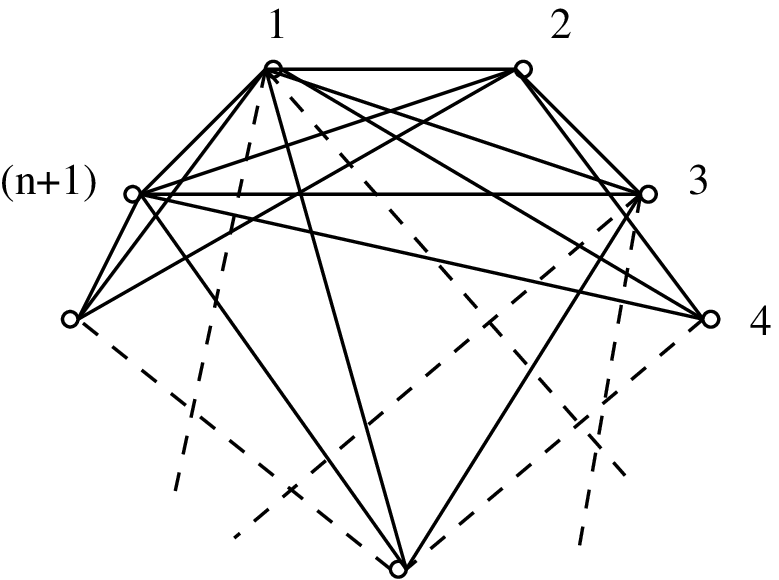} \]
Its evaluation is given by the integral

\begin{eqnarray}
K^{n+1}= \int_{\mathbf{S}^{n-1}}  dx_{1}...dx_{n+1}  \prod_{e}
K_{\rho_{e}}^{N}(r(x,y))
\end{eqnarray}
and its asymptotics which was studied in \cite{fk} and specialised to the
3-dimensional case in \cite{jm} is given by

\begin{equation}
\cos \biggl( \sum_{i<j} (V_{ij} \theta_{ij} + k \frac{\pi}{4}) \biggr)
\end{equation}
where $V_{ij}$ are volumes of $(n-2)$-simpleces and $k$ is an integer
expressed in terms of the dimension $n$, \cite{fk}. In the 3-dimensional case
we have that $V_{ij}$ are lengths of edges. The formula was obtained
in \cite{jm} and in that case, $k$ is the Hessian of the action.

All these method can be generalised to include the Lorentzian case and then
to the Newtonian limit. This is done in the following sections.

\section{Lorentzian spin networks}

For the $n$-dimensional Lorentzian case we have $n$-dimensional
Minkowski space given by $\mathbf{R}^{n}$ 
with bilinear form

\[ [x,y]= x_{0}y_{0} - x_{1}y_{1} - ... - x_{n-1}y_{n-1} \]
The unimodular group which leaves this bilinear form invariant is denoted by
$SO(n-1,1)$. 
The light cone, also known by null cone, $C$ is the set of points $x \in \mathbf{R}^{n}$ which satisfy
$[x,x]=0$. 

The subgroup of $SO(n-1,1)$ of
transformations preserving both sheets of the cone is the connected component
denoted by $S_{0}(n-1,1)$.
The language of representation theory is important in the spin foam models
formulations of quantum gravity. For our group $S_{0}(n-1,1)$, its
representations are mainly divided in a set of discrete representations 
labelled by integers and in a set of continuous representations labelled 
by real numbers. 

In this paper we restrict ourselves to the continuos representations known
as the principal unitary series. These representations are thought as 
labelling spin networks with spacelike edges. A state sum model of 
quantum gravity is constructed by considering a triangulation $\triangle$
of an $n$-dimensional manifold $M$ and considering its dual complex 
$\mathcal{J}_{\triangle}$, we construct a spin foam model by labelling each
face of $\mathcal{J}_{\triangle}$ by a principal unitary irreducible
representation of the group $SO_{0}(n-1,1)$. The state sum model is the given
by

\begin{equation}
\mathcal{Z}(M)= \int_{0}^{\infty} d\rho_{f} \prod_{f} A(f) \prod_{e} A(e)
\prod_{v} A(v) 
\end{equation}
where the integration is carried over the labels of all internal faces of the 
dual complex and $A(f), A(e), A(v)$ are the amplitudes given to the 
faces,edges and vertives of the dual complex $\mathcal{J}_{\triangle}$.
These amplitudes are given by the evaluation of certain labelled 
spin networks such as

\[ A(f)= \epsfcenter{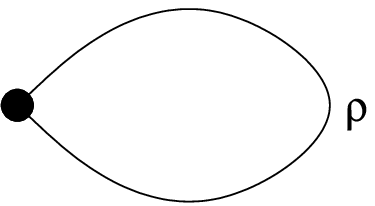} \]

\[ A(e)= 1/ \epsfcenter{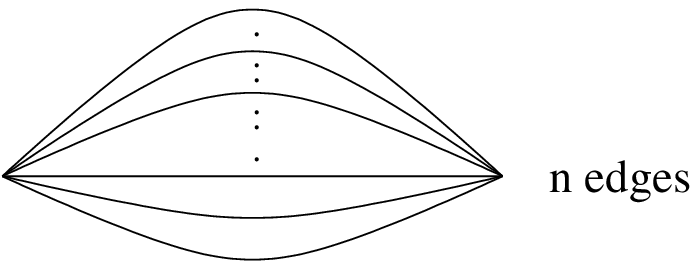} \]

\[ A(v)= \epsfcenter{lor6.eps} \]
Recall that the egdes of these graphs are labelled by principal unitary
representations of $SO_{0}(n-1,1)$. 
The vertex amplitude is given by the evaluation of the n-simplex graph.
This $n$-simplex can be seen as a collection of $n+1$ points which are all
joined together by edges.
 
The evaluation of these graphs is analogous to the Euclidean case, so the
recipe is:

\bigskip

-To each edge of the spin network we associate a propagator.
The propagators are given by zonal spherical funtions for the 
continious representations of $SO_{0}(n-1,1)$.

\bigskip

-We multiply all these propagators and integrate over a copy of an internal
space for each vertex. The internal spaces are homogeneous spaces in
which the propagators are evaluated. 

\bigskip

For the group $SO_{0}(n-1,1)$ the zonal spherical functions
for the continuos represenations are given by the Legendre functions

\begin{equation}
K_{\rho}^{L}(r)= \frac{\Gamma(\frac{n-1}{2})}
{\sqrt{\pi}\Gamma(\frac{n-2}{2})}
\int_{0}^{\pi}
( \cosh r - \cos \theta \sinh r)^{\sigma} \sin^{n-3} \theta d\theta
\end{equation}
where $\sigma = -p + i \rho$. $p$ is related to the dimension of the 
space-time as $p=(n-2)/2$, and $\rho$ labels a continuous 
representation of $SO_{0}(n-1,1)$.

Moreover, the homogeneous space is given by the $(n-1)$-dimensional
hyperbolic space $\mathbf{H}_{\infty}^{n-1}= SO_{0}(n-1,1)/SO(n-1)$.
\footnote{Let us denote the $(n-1)$-dimensional hyperbolic space simply by 
$\mathbf{H}_{\infty}$. This notation will be clear when we study the non-archimedean 
formulation.}

\subsection{The $n$-simplex}

The non-degenerate $n$-simplex is represented as the complete graph of $(n+1)$ 
vertices denoted by $K^{n+1}$.
Following the recipe for evaluation of spin networks, 
the evaluation of the $n$-simplex spin network graph would be 
given by the following integral

\begin{eqnarray}
K^{n}= \int_{\mathbf{H}_{\infty}} dx_{1}...dx_{n-1} \prod_{e}
K_{\rho_{e}}^{L}(r(x,y))
\end{eqnarray}
where we have a multiple integral over $n$ vertices of the $n$-simplex.
One of the integrals was dropped for regularisation in analogy to the
3-dimensional case \cite{jm}, and the 4-dimensional case \cite{bc}.
A problem that appears now is whether our integral (7) converges, that is,
whether it is well defined. This problem also arised in the 3-dimensional case
\cite{jm}, and in the 4-dimensional case \cite{bc}. Here we give much
evidence for its convergence in any dimension. This evidence is
very close to a proof.

\subsubsection{The convergence}

Before showing the evidence for the convergence of the $n$-simplex in any
dimension we first consider some other integral evaluations and show 
their convergence. Although this discussion follows closely the same ideas of
the 4-dimensional case \cite{bb}, it gives a generalisation to the $n$-dimensional case.

We have that for $\rho \neq 0$ our kernel 
$K_{\rho}^{L}(r)$ is well defined for small
$r$ and for large values of $r$ is asymptotic to \cite{mos} 

\begin{equation} 
K_{\rho}^{L}(r) \sim 
A_{\rho} 2^{(n-3)/2} \Gamma(\frac{n-1}{2}) e^{(1-\frac{n}{2})r}
\end{equation} 
where $n$ is the dimension of the space-time and $A_{\rho}$ is a factor which  
depends on the representation.  

\bigskip

\danger{The theta symbol and more general graphs with loops:}
We prove first that the theta symbol converges in any dimension, that is
the following evaluation converges

\begin{equation}
\epsfcenter{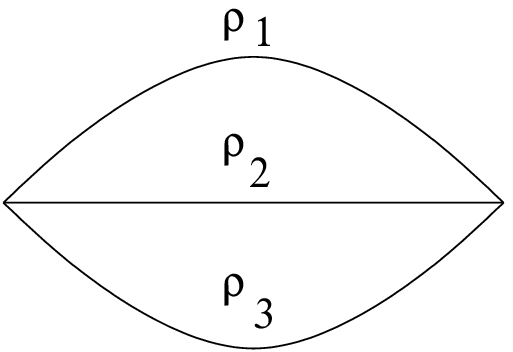}= \int_{\mathbf{H}_{\infty}} dx \ K_{\rho_{1}}(x,y) K_{\rho_{2}}(x,y)
K_{\rho_{3}}(x,y)
\end{equation} 
This follows easily from the asymptotic behaviour of our kernel where our 
integral can be written in the form

\begin{equation}
\sim A_{\rho_{1}} A_{\rho_{2}} A_{\rho_{3}}
2^{3(n-3)/2} \biggl( \Gamma(\frac{n-1}{2}) \biggr)^{3}
\int_{\mathbf{H}_{\infty}} dx \ e^{(1- n/2)r} e^{(1- n/2)r} e^{(1- n/2)r}
\end{equation} 
The volume form gives

\begin{equation}
\sim A_{\rho_{1}} A_{\rho_{2}} A_{\rho_{3}}
2^{3(n-3)/2} \biggl( \Gamma(\frac{n-1}{2}) \biggr)^{3}
\int_{}^{\infty} dr \ e^{\frac{(2- n)}{2}r} 
\end{equation}
which for $n \geq 3$ is finite. This implies that the theta symbol 
converges in any dimension greater than or equal to 3.
 
Similarly, if we consider the evaluation of the edge amplitude $A_{e}^{-1}$,
or any such graph of two vertices joined by $k \geq 3$ edges, we have that
its evaluation is given by

\begin{eqnarray}
\epsfcenter{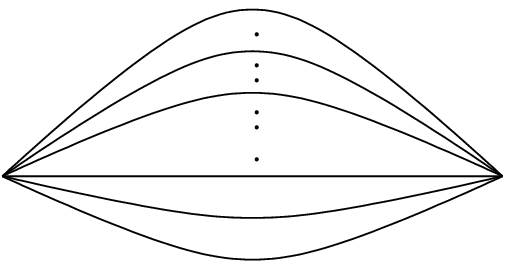} \sim A_{\rho_{1}} A_{\rho_{2}}...A_{\rho_{k}}
2^{k(n-3)/2} \biggl( \Gamma(\frac{n-1}{2}) \biggr)^{k}
\int_{}^{\infty} dr \ e^{k(1- n/2)r} e^{(n-2)r} \nonumber \\
\hspace{1.5cm} \sim A_{\rho_{1}} A_{\rho_{2}}...A_{\rho_{k}}
2^{k(n-3)/2} \biggl( \Gamma(\frac{n-1}{2}) \biggr)^{k}
\int_{}^{\infty} dr \ e^{\frac{(k-2)(2-n)}{2}r} 
\end{eqnarray}
which is finite. 

\bigskip

\danger{More general integrals:} Now we consider the following integral

\begin{equation}
I= \int_{\mathbf{H}_{\infty}} dx \ K_{\rho_{1}}(x,x_{1}) K_{\rho_{2}}(x,x_{2})...
K_{\rho_{k}}(x,x_{k})
\end{equation} 
for fixed $x_{1},...,x_{k} \in \mathbf{H}_{\infty}$. We prove that it converges.
Using the fact that our kernel $K_{\rho}(x,y)$ is asymptotic to  
$A_{\rho} 2^{(n-3)/2} \Gamma(\frac{n-1}{2}) e^{(1-\frac{n}{2})r}$
we have that our integral $I$ is asymptotic up to some factors, to

\begin{equation}
\int_{\mathbf{H}_{\infty}} dx \ e^{(1-\frac{n}{2})(r_{1}+...r_{k})}
\end{equation}
where $r_{i}=d(x,x_{i})$. In \cite{bb} it was proved that 
we can find a barycentre $b \in H^{+}$ for the points $x_{i}$, such that
\footnote{Although in \cite{bb} the case of the 3-dimensional 
hyperbolic space was considered, the proof for the existence of a 
barycentre works in any dimension} 

\begin{equation}
r:= d(x,b) \leq \frac{1}{k}(r_{1}+...+r_{k})
\end{equation}
Then in spherical coordinates around $b$ we see that our integral $I$
is bounded by

\begin{equation}
\int_{}^{\infty} e^{k(1-\frac{n}{2})r} \sinh^{(n-2)}r \ dr
\end{equation}
which converges for all $n \geq 3$ and all $k \geq 3$.

Consider again the integral (14). We have that

\begin{equation}
I \sim \int_{}^{\infty} \sinh^{(n-2)}r e^{(1-\frac{n}{2})(r_{1}+...r_{k})} \ dr
\end{equation}
Now we proceed as in \cite{bb} by breaking the integral over $r$ into two parts 
by considering:

\[ r_{1}+...+r_{k} \geq kr \ , \ M=\frac{1}{k} min_{x} (r_{1}+...+r_{k})\]
therefore we have that

\begin{eqnarray}
I \sim \int_{}^{M} e^{(1-\frac{n}{2})kM} \sinh^{(n-2)}r \ dr +
\int_{M}^{\infty} e^{(1-\frac{n}{2})kr} \sinh^{(n-2)}r \ dr \nonumber \\
\sim \frac{1}{2} \int_{}^{M} e^{(1-\frac{n}{2})kM+(n-2)r} \ dr +
\frac{1}{2} \int_{M}^{\infty} e^{(1-\frac{n}{2})kr+(n-2)r} \ dr \nonumber \\
\leq B e^{(1-\frac{n}{2})kM+(n-2)M} 
\end{eqnarray}
where $B$ is a constant factor. The triangle inequality implies that

\begin{equation}
r_{1}+...r_{k} \geq \frac{1}{k-1} \sum_{i<j}r_{ij}
\end{equation}
for all $x_{1},..,x_{k}$ where $r_{ij}=d(x_{i},x_{j})$. Therefore we have that

\begin{equation}
M \geq \frac{1}{k(k-1)} \sum_{i<j}r_{ij}
\end{equation}
which implies that

\begin{equation}
I \sim B \exp \biggl[ ((1-\frac{n}{2})k+(n-2))\frac{1}{k(k-1)}\sum_{i<j}r_{ij}) \biggr]
\end{equation}
Now we can go back to the $n$-simplex and compute for instance the integral over
$x_{n}$. This resembles an integral of the same kind as the integral $I$.
Therefore it is asymptotic to the equation (21) and we continue to integrate over the other
variables. At some step we will have to change to spheroidal coordinates as
was done in \cite{bb} for the 4-dimensional case. 
Following the procedure it can be expected that the integral that defines the
evaluation of the $n$-simplex is bounded and therefore converges.

\subsection{Asymptotics}

Once we have shown evidence for the convergence of the amplitude evaluation
of the $n$-simplex, we calculate its asymptotics.\footnote{The study of
the asymptotics of amplitudes of spin networks in quantum gravity can be found in 
many other references such as \cite{bce}, \cite{bs}, \cite{fl}, \cite{jm}}
First we notice that evaluating the integral of our kernel $K_{\rho}^{L}(r)$,
we have that 

\begin{equation}
K_{\rho}^{L}(r)= \frac{2^{(n-3)/2} \Gamma((n-1)/2)}{\sinh^{(n-3)/2} r} 
\mathfrak{B}_{\sigma +(n-3)/2}^{(3-n)/2}(\cosh r)
\end{equation} 
where $\mathfrak{B}_{\sigma +(n-3)/2}^{(3-n)/2}(\cosh r)$ are
Legendre functions.
We have that $\sigma= -p+ i \rho$ where $p=(n-2)/2$ so that

\begin{equation}
\mathfrak{B}_{\sigma +(n-3)/2}^{(3-n)/2}(\cosh r)=
\mathfrak{B}_{-(1/2)+i \rho}^{(3-n)/2}(\cosh r)
\end{equation}
For large value of $\rho$ we have that our kernel can be approximated by
the asymptotic behaviour of 
$\mathfrak{B}_{-(1/2)+i \rho}^{(3-n)/2}(\cosh r)$ 
which is given by

\begin{equation}
\sqrt{\frac{2}{\sinh r}} \rho^{(2-n)/2} 
\cos \biggl( \rho r + \frac{(3-n) \pi}{4} -
\frac{\pi}{4} \biggr)
\end{equation}
We now have that our $n$-simplex evaluation can be written as 

\begin{eqnarray}
K^{n+1}= \int_{\mathbf{H}_{\infty}} dx_{1}...dx_{n-1} \prod_{i < j}
e^{i \sum_{i < j} \epsilon_{ij} \rho_{ij} r_{ij}+ (2-n) \pi/4} 
\end{eqnarray}
The function given by $S= \sum_{i < j}\epsilon_{ij} 
\rho_{ij} r_{ij}+ (2-n) \pi/4$ is called the action.

We rescale all of our representations $\rho_{ij}$ by a common factor
$\alpha \rho_{ij}$ and look for the behaviour of our
integral formula when $\alpha \rightarrow \infty$.
We use the stationary phase method to evaluate the asymptotics of 
our integral formula (25). Moreover we restrict to the
non-degenerate configurations where all $r_{ij} \neq 0$.

Given our non-degenerate $n$-simplex, there are $(n+1)$  
timelike unit vectors $n_{i}$ which are normals to the $(n-1)-$simpleces.
There is a notion of Lorentzian angle, from which a 
Schl\"afli identity follows, \cite{bb}. This identity is given by

\begin{equation}
\sum_{i \leq j} V_{ij} d \Theta_{ij}=0
\end{equation}
Some of our normal vectors are future pointing and some are past pointing.
We know that our points $x_{i}$ live on the hyperbolic hyperspace $\mathbf{H}_{\infty}$
which implies that these vectors are all timelike and future pointing.
Then $x_{i}=a_{i}n_{i}$ where $a_{i}=1$ if $n_{i}$ is future pointing,
or $a_{i}=-1$ if $n_{i}$ is past pointing.

By taking this into account we then vary the action and constrain such
variation by a Lagrange multiplier which finally fixes the $\epsilon$'s 
up to an overall sign. 
We will then have an asymptotic behaviour given by an oscillatory function
analogous to the Euclidean one.

\section{Newtonian spin networks}

This case of Newtonian spin networks has a mathematical
sense, as it is a limit of the Euclidean and Lorentzian
cases. However, its physical interpretation as a
Newtonian quantum gravity theory is not clear to the author.
 
First of all, as it is mentioned in \cite{i}, quantum gravity
refers to the attempts to unify general relativity and quantum theory.
If gravity was nothing but the Newtonian well known static force,
the constuction of a corresponding quantum theory would be a simple
and uninteresting affair.

But even more, it may not have sense at all. This is because of the following 
reasoning: The non-degenerate scalar product which gives rise to the
inhomogeneous symmetry of space-time is interpreted physically 
as a limiting case
of the Lorentzian one when the speed of light $c \rightarrow \infty$.
This is also correct form the Newtonian theory point of view where the principle of
the universal speed limit given by the velocity of ligth $c$ is no longer true.
In Newtonian theory a body can reach any speed and so light could travel
so fast in a reference frame. Therefore $c$ is not a constant any more and 
the Planck length

\[ \ell_{P}= \biggl( \frac{Gh}{c^{3}} \biggr)^{\frac{1}{2}} \]
has no meaning at all as a constant. If light $c \rightarrow \infty$,
the Planck length $L_{P} \rightarrow 0$, and so, space-time is classical.

It is also possible that the name of Newtonian spin networks is not
appropriate. Any way, the inhomogeneous limit has sense
and we have the right to study the evaluation of its spin networks.    
 
Mathematically the inhomogeneous group $ISO(n-1)$ can be obtained by a limit
procedure from $SO_{0}(n-1,1)$ \cite{vk}. 
 
The Newtonian spin networks for $n$-dimensional quantum gravity
whose group
is $ISO(n-1)$, and whose bilinear form is degenerate
should then be evaluated by a zonal spherical function of this
group. That is,

\begin{equation}
K_{\rho}^{N}(r)= \frac{\Gamma(\frac{n-1}{2})}
{\sqrt{\pi}\Gamma(\frac{n-2}{2})}
\int_{0}^{\pi}
e^{r \rho \cos \theta} \sin^{n-3} \theta d\theta
\end{equation}
where as for the Lorentzian case, we also have a continuous set of 
representations denoted by $\rho$.

Moreover, the homogeneous space is given by 
$\mathbf{R}^{n-1}= ISO(n-1)/SO(n-1)$.

Our spherical functions (27) are Bessel functions
of the corresponding dimensions.

Therefore the evaluation of the $n$-simplex in the Newtonian framework 
is given by the integral

\begin{eqnarray}
K^{n+1}= \int_{\mathbf{R}^{n-1}} dx_{1}...dx_{n-1} \prod_{e}
K_{\rho_{e}}^{N}(r(x,y))
\end{eqnarray}
This integral seems to be well defined since for small values of 
$r$, that is when $r \rightarrow 0$ the Bessel functions approach

\begin{equation}
\frac{r^{n-3}}{2^{(n-3)} \Gamma(1+(n-3))}
\end{equation}
moreover, the asymptotics of these functions for $\rho \rightarrow \infty$ is given by

\begin{equation}
\sqrt{\frac{2}{\pi \rho r}}  
\cos \biggl( \rho r - \frac{(n-3) \pi}{2} - 
\frac{\pi}{4} \biggr)
\end{equation}
As our Lorentzian case approaches the Newtonian case for small values of
$r$, we can hope that our Newtonian tetrahedron evaluation is finite
as in our Lorentzian case.

If the evaluation of the $n$-simplex converges by a similar argument
to the Lorentzian one, then one must expect an oscillatory asymptotic behaviour.

\section{Non-Archimedean and Non-Commutative Formulations}

In this section we define two new formulations of possible spin foam models of quantum gravity
which may be related to a non-archimedean and
non-commutative property of space-time. These two models might be related
to each other as it is commented in following subsections.     
There is a new revolution of our concept of space-time coming from quantum gravity,
which is the idea that space-time is discrete
at the Planck length. Moreover, it is believed that algebra and combinatorics play an important 
role for quantum gravity and that the concept of point in space-time is non-sense.
Therefore a non-archimedean formulation includes the idea of a discreteness of space-time,
and a non-commutativity includes the idea of representing the whole information of a quantum
space-time algebraically. 

We develop this ideas here, and a more precise definition of these models require more
study.

\subsection{Non-Archimedean Spin Foam Models}

We now propose the construction of a non-archimedean 
spin foam model of quantum gravity.\footnote{It will also be called $p$-adic spin foam model} 
We see this new non-archimedean model as an alternative
way of studying covariant quantum gravity and  
the most interesting situation about this non-archimedean 
model is that it might be an alternate way of renormalisation of quantum 
gravity in the physical context of    
state sum models. Also, it gives a discreteness of space-time which is
expected at the fundamental level.

It is indeed a $p$-adic gravity constructed over a $p$-adic space-time.
The field of real numbers $\mathbf{R}$ is an extension of the
rationals $\mathbf{Q}$, but there are many other infinite 
extensions of the rationals with an equal right.
These extensions are constructed for any prime number $p$, 
and are known by the name of $p$-adic numbers $\mathbf{Q}_{p}$, 
\cite{k}, \cite{vvz}. 

As quantum field theory and string theory have their $p$-adic alternative
formulation \cite{vvz}, \cite{z}, we propose that spin foam models of quantum 
gravity have a $p$-adic formulation over a $p$-adic space-time $\mathbf{Q}_{p}^{n}$.

For instance, in the case were $n=3$ we have that the real 3-dimensional Lorentzian spin foam
model of quantum gravity and its evaluation of spin networks as studied in
\cite{jm}, has an alternative formulation as a Lorentzian $p$-adic spin foam model.

Its corresponding spin networks should be evaluated with the help of harmonic analysis
over the corresponding homogeneous spaces. These homogeneous spaces are the 
counterpart of the hyperbolic plane given by 
$SL(2, \mathbf{R})/SO(2)$. 

The corresponding $p$-adic homogeneous spaces are given by 
$PGL(2, \mathbf{Q}_{p})/PGL(2, \mathbf{Z}_{p})$ \cite{z}. 
These spaces are visualised as infinite trees with vertices of valance $p+1$.
These trees are known to be lattice hyperbolic spaces which are called Bruhat-Tits
trees(see figure 1).

\begin{figure}[h]
\begin{center}
\includegraphics[width=0.5\textwidth]{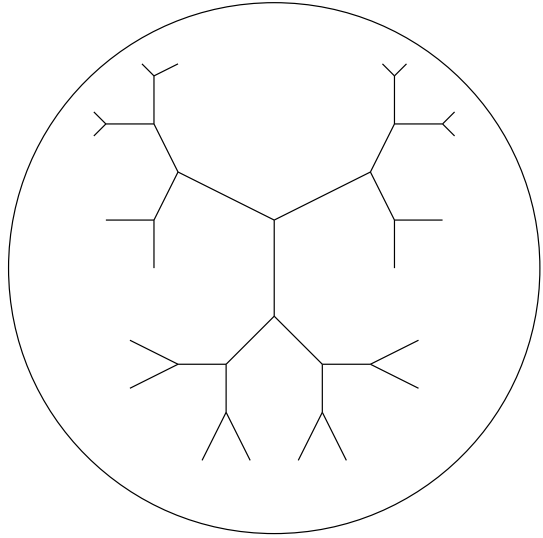}
\caption{Bruhat-Tits tree as a $p$-adic hyperbolic space. This particular figure shows 
an example of a $2$-adic hyperbolic space}
\end{center}
\end{figure}
We denote these spaces by $\mathbf{H}_{p}$. In this language the classical hyperbolic plane
is denoted by $\mathbf{H}_{\infty}$.
This means that we recover our previous notion of archimedean 
spin foam models of quantum gravity.
 
There is a notion of distance and of geodesics between vertices in these $p$-adic 
hyperbolic spaces $\mathbf{H}_{p}$. As our $p$-adic hyperbolic space is a tree, there is
a unique path of edges between any two vertices. This unique path is considered
to be a geodesic between the vertices, and the number of edges that form this path
is the distance between these vertices. Then we can choose any vertex to be
the centre of our space.

We know that in order to evaluate the amplitudes of spin networks, we need
to do some harmonic analysis on our homogeneous spaces and find the
corresponding zonal spherical functions. Fortunately, in \cite{c}, we 
can find a way to do harmonic analysis on trees. In \cite{f}, a study
of zonal spherical functions over trees is found as well as the expression
for the $p$-adic discrete laplacian. Moreover, the asymptotic behaviour of
the $p$-adic zonal spherical functions for large distances $r$ on the
$p$-adic hyperbolic space is given by

\begin{equation}
\phi_{\lambda}^{p} \sim A_{\lambda} \ p^{-\frac{r}{2}}
\end{equation}
were $A_{\lambda}$ is a factor that depends on a real parameter $\lambda$.
These $\lambda$'s should be thought as labelling edges of $p$-adic
spin networks.

\bigskip

\danger{Evaluation of $p$-adic spin networks}
We then give the recipe for the evaluation of $p$-adic spin networks,
a recipe who resembles the one of spin networks 
of archimedean spin foam models of quantum gravity, a one that we already know 
very well.

Given a spin network graph whose edges are labelled by numbers $\lambda$,
we define its evaluation as follows:

\bigskip

- To each edge of the spin network we associate a $p$-adic propagator.
This $p$-adic propagator is given by the $p$-zonal spherical
functions $\phi_{\lambda}^{p}$.

\bigskip

-We multiply all these propagators and integrate over a copy of an $p$-adic internal
space for each vertex. These $p$-adic internal spaces are given by the $p$-adic 
hyperbolic spaces $\mathbf{H}_{p}$, also known as Bruhat-Tits trees.   
  
\bigskip

It is interesting to study the possible convergence of these
$p$-adic spin foam models and propose them as a possible regularisation
of quantum gravity. These models are seen as an alternative description
of quantum gravity which in a way also resembles a kind of lattice 
formulation because of the Bruhat-Tits trees which play an important
role in the formulation. 
In this way we can see the $p$-adic formulation more than an 
alternative formulation, but as a generalization of spin foam models
which is based on a discrete space-time; more like in lattice gauge theories.
We should be able to obtain the continuous picture of our space-time,
which suggests a possible adelic formulation of quantum gravity.

We think that all these ideas require a formal formulation and careful study
and leave them to future work.

Another very related formulation to the $p$-adic one is given below
and it is seen as a non-commutative one.

\subsection{Non-Commutative Formulations}

Nowadays there has been a deep relation between quantum field
theory and noncommutative geometry. There are many issues of
the 20th century physics which have been studied in the 
noncommutative framework. 

In this section we propose a study of quantum gravity, based on a non-commutativity
of space-time based on spin
foam models. This implies
that we should have a noncommutative counterpart of
the classical BF action and of the constrained one which
gives gravity. Moreover, the Feynman path integrals can
be deformed to a noncommutative framework. This implies
that they should correspond to spin foam models of the
noncommutative BF and gravity theories.

This non-commutative picture is related in a way to the
non-archimedean one, but it is more general.

We suppose that there is such way to obtain a 
spin foam theory of a noncommutative gravity theory and    
propose a way to evaluate the corresponding spin networks.
These spin networks are completely abstract as in general 
our quantum spaces do not have points. We called such
networks by the name of $q$-spin networks.

\bigskip

\danger{$q$-spin networks}

Given a state sum model of non-commutative quantum gravity,
we need to know how to evaluate the amplitudes of the vertices, edges 
and faces networks. We propose that the evaluation is given analogously to
the case of spin networks of quantum gravity with vanishing cosmological
constant by introducing a $q$-parameter such that $0 \ < \ q \ < \ 1$.

We define the evaluation of $q$ spin networks in general. 
We now suppose we have a closed spin network and define its amplitude
as a Feynman integral which evaluation rules are given as follows:

\bigskip

-To each edge of the spin network we associate a $q$-propagator.
This $q$-propagator is given by $q$-zonal spherical
functions of the representations of a quantum group $G_{q}$.
These $q$-spherical functions are given by eigenfunctions of the
$q$-Laplace-Beltrami operator which is the Casimir element of our
quantum group.

\bigskip

-We multiply all these propagators and integrate over a copy of an $q$-internal
space for each vertex. These $q$-internal spaces are analogous to the homogeneous spaces in
which the propagators are evaluated. For the three cases of Euclidean, Lorentzian 
and Newtonian quantum gravity, these $q$-internal spaces are $q$-spheres(quantum spheres), 
$q$-hyperbolic spaces and $q$-planes(quantum planes).

\bigskip

For the Lorentzian case, there is a relation between the $p$-adic 
zonal spherical functions and the $q$-zonal sphercal functions \cite{f}.
This relation could lead us to a better understanding between the 
$p$-adic case and the non-commutative one. Moreover, this would lead us
finally to a connection with the clasical case. We do not study any of this
relations in the present paper but propose them as future problems.

It is also interesting to notice that we have another important similarity
between the classical and quantum spherical functions,
implying a possible way to formulate a meaning of non-commutative
spin foam models in a formal way. For instance, in the case
of Euclidean quantum gravity in which the lie algebra is given by $\mathbf{sl}_{2}$
the propagators which are zonal spherical functions(Legendre polynomials), 
are solutions of the well known Knizhnik-Zamolodchikov equations
\cite{kz}.
In fact these functions are special cases of the more general Gauss hypergeometric function
given by

\begin{equation} 
F(a,b,c;z)=\sum_{k=0}^{\infty} \frac{(a)_{k}(b)_{k}}{k! (c)_{k}} z^{k}
\end{equation}
which has an integral representation given by

\begin{equation}
F(a,b,c;z)= \frac{\Gamma(c)}{\Gamma(b)\Gamma(c-b)} \int_{0}^{1} t^{b-1}(1-t)^{c-b-1}
(1-tz)^{-a} dt
\end{equation}
Moreover, all solutions of the Knizhnik-Zamolodchikov equations are expressed in terms
of generalized hypergeometric functions.

For the  $q$-spherical special functions(little $q$-Legendre polynomials) 
we also have that such polynomials have similar properties as
the usual Legendre polynomials. They are also orthogonal polynomials on the interval
$[0,1]$ with respect to a measure known as $q$-beta measure. They are also eigenfunctions
of a second order $q$-differential operator.
Moreover, they are solutions of the
$q$-deformed versions of the Knizhnik-Zamolodchikov equations
which in fact are special cases of the more general $q$-deformed versions 
of the Gauss hypergeometric functions given by

\begin{equation} 
_{2}\phi_{1}(q^{a},q^{b};q^{c};q,z)=\sum_{k=0}^{\infty} 
\biggl( \prod_{j=0}^{k-1} \frac{[a+j][b+j]}{[c+j][1+j]} \biggr) z^{k}
\end{equation}
which has an integral representation given by

\begin{equation}
_{2}\phi_{1}(q^{a},q^{b};q^{c};q,z)= \frac{\Gamma_{q}(c)}{\Gamma_{q}(b) \Gamma_{q}(c-b)} 
\int_{0}^{1} t^{b-1} \frac{[(1-tz)^{-a}]}{[(1-t)^{b-c}]}
\frac{d_{q}t}{1-t}
\end{equation}
where this integral is a Jackson integral.

There is an extremely close relationship between these hypergeometric functions
and representation theory of Kac-Moody Lie algebras and quantum groups. 

Similarly we can generalise all this to the Lorentzian and Newtonian cases,
where we have $q$-Legendre functions 
and $q$-Bessel 
functions respectively. 

For instance there are two types of $q$-Bessel functions in equall foot as generalising
the classical Bessel function. Their expressions are given by \cite{vk2}

\begin{equation}
J_{\nu}^{1}(r;q)= \frac{1}{(q;q)_{\nu}} \biggl( \frac{r}{2} \biggr)^{\nu} _{2}\phi_{1}
\biggl( 0,0;q^{\nu +1};q,-\frac{r^{2}}{4}  \biggr)
\end{equation}

\begin{equation}
J_{\nu}^{2}(r;q)= \frac{1}{(q;q)_{\nu}} \biggl( \frac{r}{2} \biggr)^{\nu} _{0}\phi_{1}
\biggl( q^{\nu +1};q,-\frac{r^{2}q^{\nu +1}}{4}  \biggr)
\end{equation}
Both these expressions are related by a factor and then up to these factor
we can think of them as our propagators.

\bigskip

\section{Conclusions}

We gave much evidence for the possible convergence of the $n$-simplex
in $n$-dimensional Lorentzian quantum gravity. We think that a similar
reasoning would follow in the limit of Newtonian spin networks. This
Newtonian case may need a good physical interpretation, as well as
a argument of its importance to spin foasm models of quantum gravity.
In \cite{bce}, the study of degenerate configurations of the $10-j$ symbol 
in Euclidean and Lorentzian 4-dimensional 
quantum gravity, led the author to
conjecture that its evaluation is related to the evaluation  
of the so called degenerate spin networks which in our language
are given by Newtonian spin networks. It is then interesting
to understand better this limit of Newtonian spin networks
and of its importance to spin foam models.

We have also proposed that a non-archimedean and non-commutative formulations of
quantum gravity in terms of spin foam models may be possible. There is however
much work to do, so that a formalisation to these ideas should be given.  

We could try to describe both theories in terms of partition functions
of a kind of constrained $BF$-theory. The corresponding lagrangians may be 
$p$-adic and non-commutative respectively.

The thought is that these models may give a regularisation of
quantum gravity and that in the $p$-adic case, a 
discreteness is already implicit in the
space-time. Even more, an adelic continuation of this formulation
may be interesting which may include the already known spin foam models.

\vspace{2.0cm}

\danger{Acknowledgments}

I want to thank Laurent Freidel for reading the manuscript and
for his suggestions on a mutual correspondence. I also thank to Centro de
Investigaci\'on en Matem\'aticas(CIMAT) for all the support the centre has
given me.

\end{document}